# Mid-infrared laser chaos lidar

Kai-Li Lin,[1] Peng-Lei Wang,[1] Yi-Bo Peng,[1] Shiyu Hu,[2] Chunfang Cao,[3] Cheng-Ting Lee,[4] Qian Gong,[3] Fan-Yi Lin,[4] Wenxiang Huang,[2,*] and Cheng Wang,[1,5,*]

[1]*School of Information Science and Technology, ShanghaiTech University, Shanghai 201210, China*
[2]*Raytron Research, Raytron Technology CO., LTD, Huaxiu RD 66, Wuxi 214000, China*
[3]*Key Laboratory of Terahertz Solid State Technology, Shanghai Institute of Microsystem and Information Technology, Chinese Academy of Sciences, Shanghai 200050, China*
[4]*Institute of Photonics Technologies, Department of Electrical Engineering, National Tsing Hua University, Hsinchu 30013, Taiwan*
[5]*Shanghai Engineering Research Center of Energy Efficient and Custom AI IC, ShanghaiTech University, Shanghai 201210, China*
*\*wangcheng1@shanghaitech.edu.cn; wenxiang.huang@raytrontek.com*

**Chaos lidars detect targets through the cross-correlation between the back-scattered chaos signal from the target and the local reference one. Chaos lidars have excellent anti-jamming and anti-interference capabilities, owing to the random nature of chaotic oscillations. However, most chaos lidars operate in the near-infrared spectral regime, where the atmospheric attenuation is significant. Here we show a mid-infrared chaos lidar, which is suitable for long-reach ranging and imaging applications within the low-loss transmission window of the atmosphere. The proof-of-concept mid-infrared chaos lidar utilizes an interband cascade laser with optical feedback as the laser chaos source. Experimental results reveal that the chaos lidar achieves an accuracy better than 0.9 cm and a precision better than 0.3 cm for ranging distances up to 300 cm. In addition, it is found that a minimum signal-to-noise ratio of only 1 dB is required to sustain both sub-cm accuracy and sub-cm precision. This work paves the way for developing remote chaos lidar systems in the mid-infrared spectral regime.**

## 1. INTRODUCTION

Commercial light detection and ranging (LIDAR) systems typically use pulsed semiconductor lasers as light sources, with the ranging distance obtained from the time-of-flight (TOF) of the optical pulse back-scattered from targets [1,2]. In comparison with pulsed lidars, frequency-modulated continuous-wave (FMCW) lidars utilize frequency-chirping lasers as light sources, with the ranging distance extracted from the beat frequency between the back-scattered signal and the local reference one [3,4]. FMCW lidars are capable of obtaining both the distance information and the velocity information simultaneously. However, developing narrow-linewidth laser sources with highly linear chirping remains technically challenging [5]. When multiple pulsed or FMCW lidars operate simultaneously, they are susceptible to the interference and the jamming effects, due to the similarity of regular pulsed or chirping waveforms [6-8]. In order to mitigate these effects, random-modulated continuous-wave (RMCW) lidars have been introduced, where the laser source was modulated by a pseudo-random binary sequence (PRBS) [9,10]. Nevertheless, the RMCW lidars require high-speed driven electronics and optical modulators to modulate the laser source, which inevitably raises the cost of the lidar system.

Chaos lidars leverage laser chaos as a light source, which can eliminate the requirement of high-speed electronics and thereby reduce the system cost [11]. Laser chaos is a series of irregular pulse trains, which are usually produced from a deterministic laser system with external perturbation [12]. For instance, semiconductor lasers subject to optical feedback or optical injection can easily produce broadband chaos without the need for high-speed modulation [12,13]. Similar to RMCW lidars, chaos lidars extract the TOF from the cross-correlation between the echo chaos signal from the target and the local reference one. Owing to the random nature of chaotic pulse trains, chaos lidars exhibit strong immunity to the interference and the jamming effects [14,15]. The concept of chaos lidar was firstly proposed and demonstrated by Lin and Liu in 2004 [11]. In recent years, National Tsing Hua University has made substantial contributions in the development of chaos lidar systems. To name a few, Cheng *et al.* demonstrated the first 3D chaos lidar with a detection range of up to 100 m in the year of 2018 [16]. In addition, both the accuracy and the precision of the lidar system reached down to the sub-centimeter range. In 2022, the field of view of the chaos lidar was improved up to 24.5°×11.5° by utilizing quadrant avalanche photodiodes [17]. In order to improve the signal-to-noise ratio (SNR) of chaos lidar systems, various schemes have been employed to increase the pulse energy of chaos pulse trains, such as the time gating technique [16], the gain switching technique [18-20], and the pulsed master oscillator power amplifier technique [21]. In addition to the chaos generation from semiconductor lasers with external perturbation, Chen *et al.* proposed to produce chaos using a microring resonator pumped by a laser source [22]. The chaos generation is attributed to the Kerr nonlinearity and the thermo-optical effect. Interestingly, different channels of chaos are generated in parallel at different longitudinal modes, and these channels are mutually orthogonal. This microring chaos source enables the achievement of 3D parallel chaos lidar with only one laser source, instead of multiple ones [22,23].

Most chaos lidars discussed in the above section are operated in the near-infrared C-band (1530-1565 nm) spectral regime. However, the mid-wave infrared regime (MWIR, 3-5 $\mu$m) is the low-loss transmission window of the atmosphere. The transmission attenuation of the MWIR light is only about 60% of that of C-band light, benefiting from the lower scattering effect [24,25]. In addition, the mid-infrared light demonstrates strong resistance to the turbulence effect of the atmosphere [26,27]. The above advantages have facilitated the successful demonstration of mid-infrared free space optical communication systems in recent years [25,27-29]. For a given transmitter power and communication distance, the lower propagation loss and the weaker turbulence effect of the mid-infrared light lead to a higher receiver power and hence a lower bit error rate of the communication link. Similarly, mid-infrared lidars benefit from these advantages, enabling a higher SNR and/or a longer detection range compared to their near-infrared counterparts [30]. In order to develop mid-infrared chaos lidar systems, proper laser sources emitting in the mid-infrared regime are essential. Interband cascade lasers (ICLs) are power-efficient mid-infrared semiconductor laser sources, with optimal performance in the spectral range of 3 to 6 $\mu$m [31-36]. Besides, the lasing wavelength of ICLs is extendable up to more than 10 $\mu$m [37,38]. Similar as common near-infrared semiconductor lasers, ICLs belong to the category of class-B lasers as well, where the carrier lifetime (around 1 ns) is significantly longer than the photon lifetime [12,39]. We have successfully demonstrated the generation of broadband chaos from ICLs, using the perturbation of optical feedback and optical injection, respectively [40,41]. Our recent work proved that the chaos bandwidth of ICLs was as high as 6 GHz, which was comparable to that of near-infrared counterparts [42]. In this work, we demonstrate a proof-of-concept mid-infrared chaos lidar, by employing an ICL with optical feedback as the laser chaos source. This mid-infrared chaos lidar, for the first time, achieves sub-centimeter accuracy and sub-centimeter precision for detection targets ranging up to 300 cm. Both the SNR and the ranging distance of the chaos lidar can be further improved through utilizing a high-power ICL and/or highly sensitive avalanche photodiodes.

## 2. EXPERIMENTAL SETUP AND RESULTS

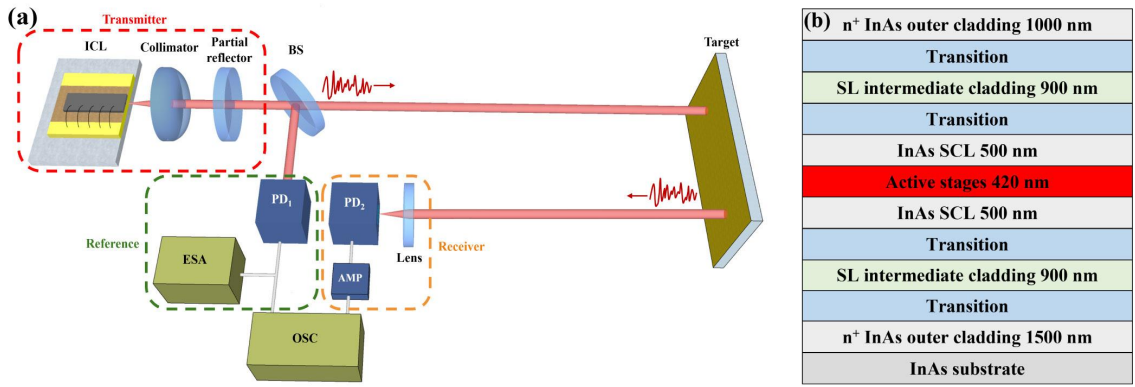

**Fig. 1.** (a) Experimental setup of the mid-infrared chaos lidar. BS: beamsplitter; PD: photodetector; AMP: amplifier; ESA: electrical spectrum analyzer; OSC: oscilloscope. (b) Epitaxial layer structure of the InAs-based ICL.

Figure 1(a) illustrates the schematic structure of the mid-infrared chaos lidar system. The system comprises three building blocks: the laser chaos transmitter, the signal receiver, and the local reference module. In the transmitter module, a Fabry-Perot ICL acts as the mid-infrared laser source. The ICL is pumped by a DC current source (Newport, LDC-3736), and the operation temperature is maintained at 20 °C by a thermo-electric cooler. The laser is collimated by an aspheric lens with a focal length of 4.0 mm. The optical feedback is provided by a partial reflector with a reflectivity of 50%, corresponding to a feedback ratio of -3 dB, defined as the ratio of the reflected light power to the laser emission power. The reflector is located about 50 cm away from the laser, resulting in an external cavity frequency of about 300 MHz. The optical feedback triggers the generation of broadband chaos from the ICL. The chaos light is split into two branches by a beam splitter (BS, 35%:65%). 35% of the light goes to the local reference module. The reference signal is detected by a HgCdTe photodetector (PD$_1$, Vigo) with a detection bandwidth of 600 MHz. The electrical spectrum of the chaos signal is measured by an electrical spectrum analyzer (ESA, Keysight N9040B) with a bandwidth of 50 GHz and a resolution of 500 kHz. Meanwhile, the temporal waveform is recorded on a digital oscilloscope (OSC, Keysight DSAZ594A) with a sampling rate of 80 GSample/s. The other 65% branch of chaos light is directed toward the target, and the echo signal is collected by a focus lens with a diameter of 71 mm and a focal length of 50 mm in the receiver module. The echo signal is detected by another HgCdTe photodetector (PD$_2$, Vigo) with a bandwidth of 560 MHz. The electrical echo signal is amplified by using a low-noise amplifier (AMP, Pasternack) with a gain of 32 dB and a cutoff frequency of 3 GHz. The corresponding temporal waveform is subsequently recorded on another channel of the OSC. The optical spectrum of the ICL is measured by a Fourier transform infrared spectrometer (FTIR, Bruker Vertex 80) with a resolution of 0.08 /cm. The targets used in the experiment are a mirror reflector with a reflectivity of 98% (Thorlabs, PF05-03-M01) and a diffuse reflector with a grit of 1500 (Thorlabs, DG10-1500-M01), respectively. Each target range measurement is repeated 100 times, and the mean value is recorded. The ICL employed in the experiment was grown on a heavily n-doped InAs substrate using molecular beam epitaxy (MBE) [43]. Although most commercial ICLs are grown on the GaSb substrate, InAs-based ICLs are able to emit at longer wavelengths. In addition, InAs-based ICLs exhibit better thermal conductivity and higher optical confinement factor of the active region [44,45]. Figure 1(b) illustrates the epitaxial layer structure of the InAs-based ICL. The active region consists of ten cascading stages of standard W-shaped type-II quantum wells [46]. It is enclosed by two 500-nm InAs separate confinement layers (SCLs). The cladding layers are composed of a 900-nm InAs/AlSb superlattice (SL) intermediate cladding and an n$^+$-doped InAs outer cladding. The thicknesses of the top and bottom outer claddings are 1000 and 1500 nm, respectively. Several transition layers are incorporated to mitigate the parasitic voltage caused by the large conduction band discontinuity between adjacent regions. These transition regions consist of graded InAs/AlSb quantum wells, which are doped at the same level with the InAs/AlSb SL claddings. After MBE growth, the wafer was processed into laser cavities with a length of 2 mm and a ridge width of 10 $\mu$m. Meanwhile, both facets of the lasers were uncoated. The ICL was mounted epitaxial-side-down on a copper heat sink.

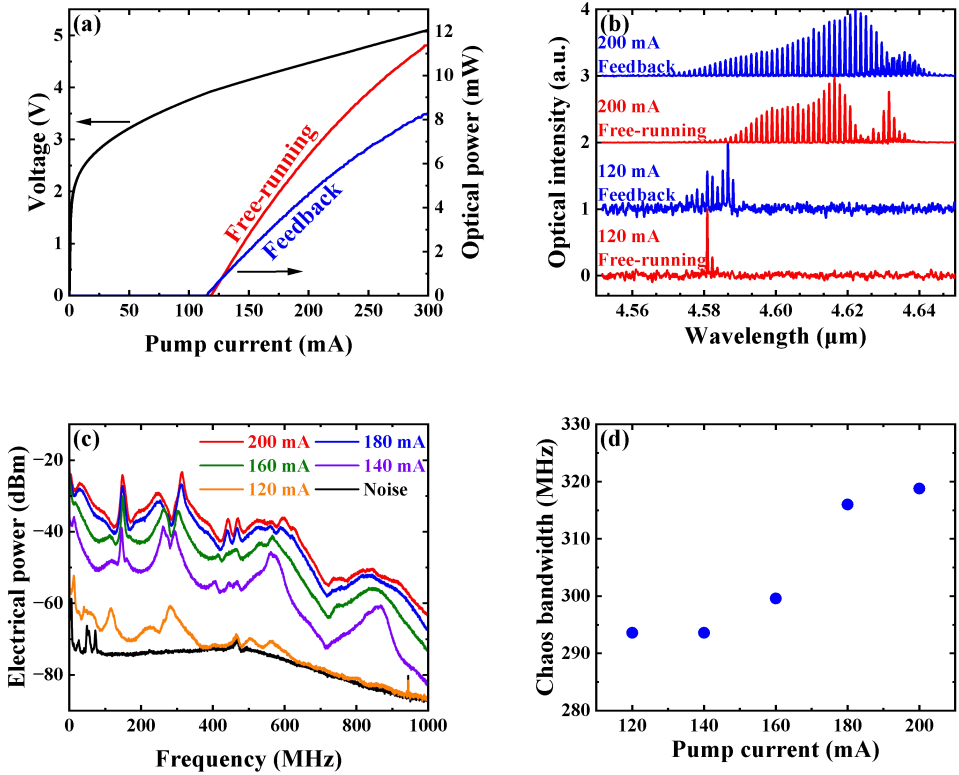

**Fig. 2.** (a) L-I-V curves and (b) optical spectra of the ICL with and without optical feedback. (c) Chaos spectra and (d) chaos bandwidth for various pump currents.

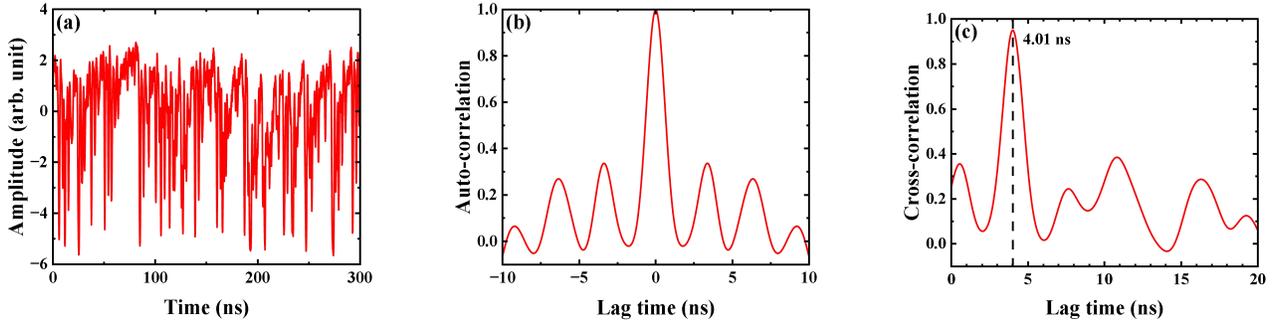

**Fig. 3.** (a) Temporal waveform, (b) autocorrelation, and (c) cross-correlation of the ICL chaos. The cross-correlation peak at 4.01 ns in (c) translates into a target distance of 60.1 cm.

Figure 2(a) shows that the threshold current of the free-running ICL is $I_{th}$ = 119 mA, and the maximum output power is more than 11 mW. The threshold voltage of the ICL is 3.9 V, and the series resistance extracted from the slope of the above-threshold voltage is 6.4 Ω. The emission wavelength of the free-running ICL pumped at 120 mA in Fig. 2(b) is around 4.58 μm. When the pump current increases to 200 mA, the lasing wavelength red shifts to be around 4.62 μm due to the thermal effect. When applying optical feedback with the feedback ratio of -3 dB, the threshold of the laser in Fig. 2(a) reduces slightly down to 116 mA. The output power of the laser with feedback is higher than the free-running case for pump currents up to 127 mA, but becomes lower for currents above 127 mA. Meanwhile, the optical feedback leads to more modes lasing in Fig. 2(b), owing to the reduction of the lasing threshold [12]. The ICL with optical feedback produces broadband chaos at various pump currents. Figure 2(c) shows that the electrical spectrum of chaos is substantially higher than the background noise. The electrical spectrum shows multiple peaks around 300 MHz as well as its harmonic frequencies. This is due to the existence of external cavity modes, and the external cavity frequency of 300 MHz is given by the reciprocal of the round-trip time of the external cavity [12].

The bandwidth of the chaos is defined as the frequency range between the DC and the cutoff frequency, where the electrical power is 80% of the total power [47]. Figure 2(d) shows that the chaos bandwidth generally rises with the pump current from 294 MHz at 120 mA up to 319 MHz at 200 mA. However, it is remarked that the measurement of the chaos spectrum is likely to be limited by the photodetector (600 MHz). Therefore, the measured chaos bandwidth in Fig. 2(d) is determined by both the resonance frequency of the ICL and the detection bandwidth of the photodetector, rather than solely by the former [42].

In the following experiment, the pump current of the ICL is fixed at 200 mA. The output power of the free-running laser is 6.4 mW, and the power of the laser with optical feedback decreases to 4.6 mW. Therefore, the average power of the laser chaos sending to the target branch in Fig. 1(a) is 1.4 mW and the one going to the reference is 0.8 mW. Figure 3(a) illustrates an example of the temporal waveform of the ICL chaos, which exhibits typical irregular behavior with random fluctuations. Figure 3(b) shows the autocorrelation of the chaos series in Fig. 3(a), which exhibits multiple sidelobes. These sidelobes are known as the time-delay signature [48].

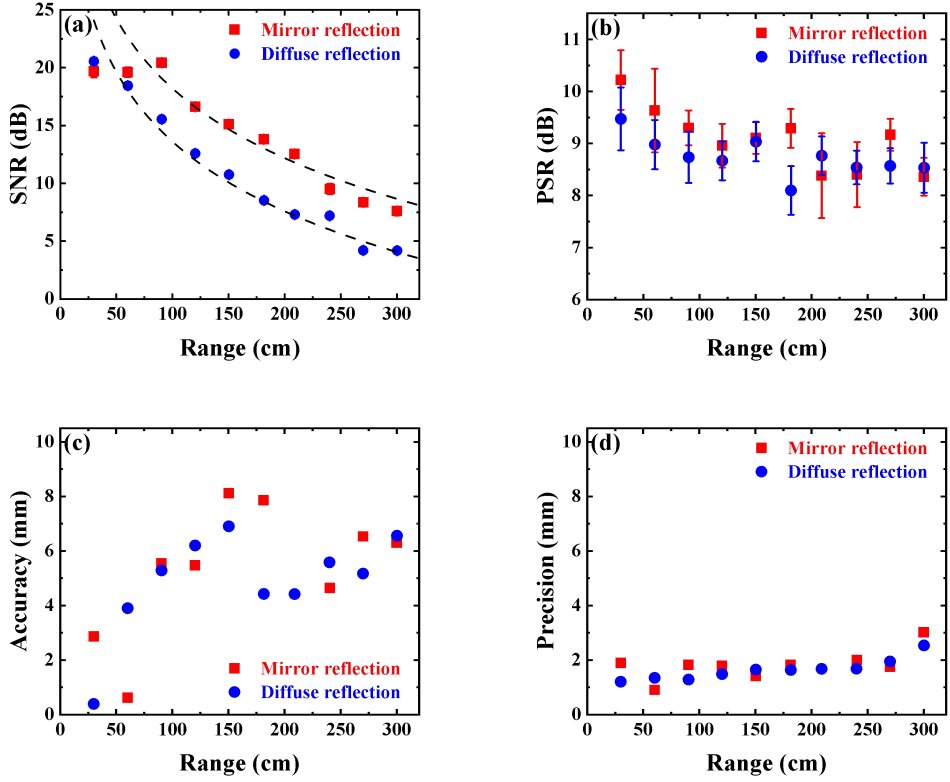

**Fig. 4.** (a) SNR, (b) PSR, (c) accuracy, and (d) precision of the mid-infrared chaos lidar versus the target range. Squares denote the mirror reflection and dots denote the diffuse reflection. The error bar stands for the standard deviation of the repeated measurements. The dash lines in (a) are least-squares fitting curves of the measured results using the reciprocal of quadratic function.

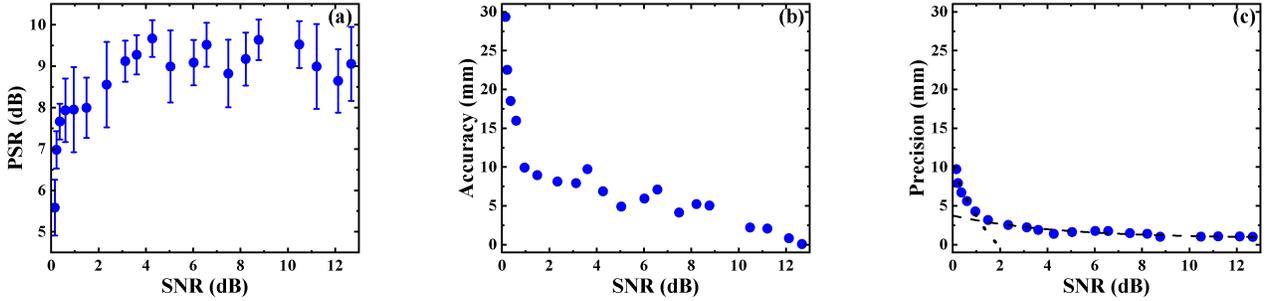

**Fig. 5.** Impact of the SNR on (a) the PSR, (b) the accuracy, and (c) the precision of the mid-infrared chaos lidar. The target distance is 150.3 cm, and the reflection is diffuse reflection. The dot line in (c) denotes the least-squares fitting using the reciprocal of the quadratic function, and the dash line denotes the fitting using the reciprocal function.

The interval between adjacent lobes corresponds to the round-trip delay time of the optical feedback. This interval of 3.4 ns suggests that the distance between the ICL and the reflector in Fig. 1(a) is precisely 51.0 cm. This time-delay signature can be suppressed by using double optical feedback [49] or distributed optical feedback [50] in future work. The target distance of the chaos lidar is extracted from the cross-correlation between the echo signal $S_R(t)$ from the target and the local reference signal $S_T(t)$, which is expressed as

$$\rho(\tau) = \frac{\langle [S_T(t) - \langle S_T(t) \rangle][S_R(t+\tau) - \langle S_R(t) \rangle] \rangle}{\sqrt{\langle |S_T(t) - \langle S_T(t) \rangle|^2 \rangle} \sqrt{\langle |S_R(t) - \langle S_R(t) \rangle|^2 \rangle}} \quad (1)$$

where $\tau$ is the time lag between the echo signal and the reference one, and $\langle \cdot \rangle$ denotes the time average. In the experiment, the span of the chaos series used for every range detection is set at 2.0 μs, that is, the cross-correlation length of the chaos signals is 2.0 μs. Figure 3(c) presents an example of the cross-correlation trace with a peak at the lag time of 4.01 ns, which translates to a target distance of 60.1 cm. Similar as the case of autocorrelation in Fig. 3(b), sidelobes also appear in the cross-correlation trace due to the existence of the time-delay signature.

Figure 4(a) presents the SNR of the lidar system as a function of the target range. The SNR is defined as the ratio between the root mean square of the echo chaos signal to the root mean square of the background noise. It is shown that the SNR generally declines nonlinearly with increasing target range both for the mirror reflection (squares) and for the diffuse reflection (dots). For the mirror reflection, the SNR decreases from 19.7 dB at the range of $L = 29.9$ cm down to 7.6 dB at $L = 300.1$ cm. In contrast, the SNR for the diffuse reflection declines from 20.6 dB at the range of $L = 29.9$ cm down to 4.2 dB at $L = 300.3$ cm. As expectation, the SNR of the diffuse reflection is generally smaller than that of the mirror reflection, due to the lower reflected power. The maximum measured range of 300 cm is limited by the available space of the experimental setup. The least-squares fitting (dashed curves) of the measured results demonstrate that the SNR is inversely proportional to the square of the target range. This observation is in good agreement with the inverse-square law of the classical lidar equation [51,52]. Figure 4(b) introduces the peak-to-sidelobe ratio (PSR) to quantify the effect of the sidelobes in the cross-correlation trace (see Fig. 3(c)). The PSR is defined as the ratio of the cross-correlation peak to three times the standard deviation of sidelobes and noise floor of the cross-correlation trace [16]. Therefore, a higher PSR is favorable for easier identification of the cross-correlation peak. Figure 4(b) shows that the PSR generally reduces with increasing range for target distances below 120 cm, which is attributed to the power reduction of the echo chaos signal. In addition, the PSR of the mirror reflection is larger than that of the diffuse reflection due to higher SNR. For target distances above 120 cm, the PSR does not exhibit an obvious trend and fluctuates around 8.7 dB. This is likely due to the misalignment of the optical path from the target reflection to the photodetector. Besides, there is no significant difference in the PSR between the mirror reflection and the diffuse one. It is remarked that the trend of the PSR in Fig. 4(b) is different to that of the SNR in Fig. 4(a). This is because the sidelobes of the cross-correlation (see Fig. 3(c)) are strong, and the level of the sidelobes declines with increasing range. In case the sidelobes (or the time-delay signature) are effectively suppressed, the PSR will become highly correlated with the SNR [17]. It is known that both the SNR and the PSR affect the accuracy and precision of chaos lidars [15-17,53]. The accuracy is defined as the absolute difference between the measurement value and the actual one. The actual distance in this work refers to the one obtained from a commercial range finder with an accuracy of 2 mm (UNI-T, LM100). The measured accuracy in Fig. 4(c) generally rises with the increasing target range up to 150 cm. Beyond this range distance, the accuracy no longer increases. This tendency is in agreement with the variation of the PSR in Fig. 4(b). The accuracy for all the measured target ranges within 300 cm is below 8.1 mm. This accuracy performance is comparable to that of the near-infrared counterparts [14,16]. The precision in Fig. 4(d) characterizes the standard deviation of the repeated measurements (100 times) for each target range in the mid-infrared chaos lidar. It is found that the precision only slightly increases within the measured distances between 30 and 300 cm. Specifically, the precision for the mirror reflection rises from 1.9 mm to 3.0 mm, while the precision for the diffuse reflection increases from 1.2 mm to 2.5 mm. Therefore, this mid-infrared chaos lidar exhibits a very good precision performance with a maximum value of 3.0 mm for ranging distances up to 300 cm.

Figure 5 investigates the effect of the SNR on the mid-infrared chaos lidar performance, where the target is the diffuse reflector located at a distance of 150.3 cm. The SNR is varied through inserting a polarizer right after the beam splitter in the experimental setup of Fig. 1(a). Because the ICL emission is polarized in the transverse electric direction, the average power of the laser chaos transmitter is tuned through rotating the polarizer. Figure 5(a) unveils that the PSR initially increases rapidly with SNR, rising from 5.6 dB at SNR of 0.15 dB up to 9.7 dB at SNR of 4.3 dB. Nevertheless, the PSR saturates around 9.1 dB for SNRs larger than 4.3 dB. This observation is in agreement with the results in Fig. 4(a) and Fig. 4(b). The saturation of the PSR is likely due to the strong time-delay signature, and this saturation can be mitigated through suppressing the external cavity modes. In such way, the PSR becomes almost linearly proportional to the SNR and remains higher than the latter because of the noise filtering effect in the cross-correlation process [17]. The accuracy of the chaos lidar in Fig. 5(b) first drops sharply from 29.3 mm at SNR of 0.15 dB down to 9.9 mm at SNR of 0.95 dB. For SNRs above 0.95 dB, the accuracy declines with a much smaller slope down to 0.1 mm at SNR of 12.7 dB. For weak echo signals, Fig. 5(c) shows that the precision of the chaos lidar reduces quickly with increasing SNR from 9.7 mm at SNR of 0.15 dB down to 4.3 mm at SNR of 0.95 dB. The least-squares curve fitting (dot line) proves that the precision of the lidar is inversely proportional to the quadratic SNR in this weak signal regime [23,54,55]. For higher SNRs above 0.95 dB, the precision declines slowly with the SNR, and reaches the minimum value of 1.0 mm at SNR of 12.7 dB. On the other hand, the least-squares curve fitting (dash line) demonstrates that the precision of the chaos lidar is inversely proportional to the SNR in this strong signal regime, which is in good agreement with the basic lidar theory [16,23,56].

## 3. CONCLUSION

In summary, we have demonstrated the first laser chaos lidar in the mid-infrared spectral regime, to the best of our knowledge. The mid-infrared laser chaos source is realized through an ICL with the perturbation of optical feedback. The maximum chaos bandwidth is 319 MHz, which is limited by both the resonance frequency of the ICL and the detection bandwidth of the photodetector. The mid-infrared chaos lidar successfully achieves a sub-cm accuracy (< 0.9 cm) and a sub-cm precision (< 0.3 cm) for ranging distances up to 300 cm. Furthermore, it is found that the sub-cm accuracy together with sub-cm precision can be achieved with a minimum SNR of only 1 dB. The ranging distance of the chaos lidar can be further enhanced through using a high-power ICL source associated with highly sensitive avalanche photodiodes. In addition, we will develop remote mid-infrared laser chaos lidar systems with the 3D imaging function in future work.

**Funding.** National Natural Science Foundation of China (62475152). Science and Technology Commission of Shanghai Municipality (24TS1401500, 24JD1402400).

**Disclosures.** The authors declare no conflicts of interest.

**Data availability.** Data underlying the results presented in this paper are not publicly available at this time but may be obtained from the authors upon reasonable request.